# Computation of Reducts Using Topology and Measure of Significance of Attributes

P.G. JansiRani and R.Bhaskaran

**Abstract**—Data generated in the fields of science, technology, business and in many other fields of research are increasing in an exponential rate. The way to extract knowledge from a huge set of data is a challenging task. This paper aims to propose a hybrid and viable method to deal with an information system in data mining, using topological techniques and the significance of the attributes measured using rough set theory, to compute the reduct, This will reduce the randomness in the process of elimination of redundant attributes, which, in turn, will reduce the complexity of the computation of reducts of an information system where a large amount of data have to be processed.

**Key terms** — Data mining, topology, rough set theory, reducts.

———————————— ◆ ————————————

## 1 INTRODUCTION

FOR centuries, extracting information from data has been done manually. But extracting useful information from a huge set of data, though highly inevitable, has been highly complicating and challenging. The increasing volume and complexity of data have necessitated the application tools of automatic data processing, and methods and models for converting data into useful information and knowledge. The process of data mining, in the modern world, plays a vital role. Neural networks and traditional statistical techniques have been compared in [14], in the prediction of academic performance of a specific business school in India. Broad based study can be done in this process using rough set theory. The Rough set theory plays a significant role in reducing the complexity of the process and automatically extracting useful information from a huge set of data.

The paper is organized as follows: Section 1 deals with Introduction, Section 2 Literature review, Section 3 presents the theoretical background and the basic concepts of rough set theory and topology and Section 4 describes the methodology to compute the reduct through two algorithms: one for the computation of base using indiscernity matrix and the other to confirm the redundancy of an attribute for the given information system. Section 5 illustrates our method with an example. It concludes in section 6 dealing with the possibility for future work.

## 2 LITERATURE REVIEW

An association rule mining algorithm that can handle different types of data based on fuzzy techniques has been developed in [21]. A procedure [25] using feature selection method and rough set classifier to predict firms' revenue growth rate has been proposed. Analysis on HR data in [8], applying data mining techniques is an efficient and viable method in the decision making process. The method proposed to examine the vital factors influencing the sharing of the electronic information of the local agency has been analyzed in [1] and a model has been proposed in [2] for effective sharing of information system that plays a significant role and can be extended globally to study the process of attribute reduction. Variable precision dominance-based-rough set approach to attribute reduction that has been proposed in [12], is effective only when the conditional attributes and the decision attributes are ordinal and monotonically related. In [4], weighted rules by using a weighted rough set based method have been extracted. In [1], rough sets have been applied to identify the set of significant symptoms causing rhinology and throatology diseases and to extract decision rules in Taiwan's Otolaryngology clinic data but the scope is limited to data of rhinology and throatology only. The essential principles and methods of rough set theory for the mining classification rules to the differentiation of symptoms and signs of Traditional Chinese Medicine (TCM) for cancer patients have been introduced in [15]. There is a scope for better accuracy in the algorithm proposed to select significant features based on correlation-based feature selection and by integrating feature selection techniques used in [18] [6] [19] [3]. Balanced information gain with a novel extension, to measure the contribution of each feature is analyzed in [23]. Predominant correlation and a fast filter method which can identify relevant features as well as redundancy among relevant features without pair wise correlation analysis are studied in [11]. A new filter approach to the feature selection that uses a correlation-based heuristic to evaluate the worth of feature subsets suggested in [13], is more effective only if the features are linearly correlated. Topological structures have been used to obtain discernibility matrix and discernibility function for knowledge reduction and decision making in [10]. But in this process the elimination of the redundant attribute, in a random

- P.G.JansiRani is with Department of Mathematics, Sethu Institute of Technology, Virudhunagar-626115, Tamil Nadu, India.
- R.Bhaskaran is with School of Mathematics, Madurai Kamaraj University, Madurai-625021, Tamil Nadu, India.

manner, makes the process uncertain. A reduct construction method in [24], based on discernibility matrix simplification has been analyzed. To find all reducts based on indiscernibility-discernibility and to reduce the discernity function to a function of disjunction of conjunction discussed in [20] [17] [26] is NP hard and hence, it is necessary to find an improved method to make the process simpler.

This paper aims to propose a hybrid method to compute the reduct of an information system by reviewing the topological techniques applied in the information system in [10] and the feature selection method proposed in [18]. We find that the hybrid method proposed in this study is a viable approach in finding the minimal reduct of an information system. The process is divided into two stages: (i) Ranking the features according to the significance of the attributes, which is taken as a probable predicting ability of attributes, and (ii) confirmation of the redundancy of the ranked attributes by applying the algebraic topological techniques in the information system.

## 3 TEHEORETICAL BACKGROUNDS

### 3.1 Basic Concepts of Topology

As demonstrated in this document, the numbering for sections upper case Arabic numerals, then upper case Arabic numerals, separated by periods. Initial paragraphs after the section title are not indented. Only the initial, introductory paragraph has a drop cap.

Topology branch of mathematics deals with the concept of "Unaffected by changing the size, shape and dimension". Algebraic Topology includes the concept of set theory in Topology. A topology on a set X is a collection of $\Im$ subsets of X which has the following properties, $\Phi$ and X are in $\Im$.
(i)The union of the elements of any sub collection of $\Im$ is in $\Im$.
(ii)The intersection of the elements of any finite sub collection of $\Im$ is in $\Im$.
A set X for which a topology $\Im$ has been specified is called a topological space.

A linearly independent subset S of the vector space L which span the whole space L is called a basis of L.

### 3.2 Base for a Topology

Given a Topological space (X, $\Im$) a collection B (X) of open subsets of X is known to be a base for topology $\Im$ if B(X) $\subset \Im$

Each member of $\Im$ can be expressed as the union of the member of B(X)

### 3.3 Sub-base

If (X, $\Im$) topological space, a collection A of subsets of X is said to be a sub-base for T iff finite intersections of members of A form a base for $\Im$.

If (X, $\Im$) topological space and a collection A is a sub-base of X then
base <A> = base< base< A1 U A2 >

= base< base< A1> U base< A2> >
base <A> = base<G1 U G2>, where A= A1U A2, G1= base< A1>, G2=<base< A2>. It can also be generalized as base <A> = base< A1 U A2 …U An > = base< base< A1> U base< A2>…U base<An> > base <A> = base< A1U A2 … U An > = base<G1 U G2 … UGn> where A= A1UA2 … U An, G1= base< A1>, G2=<base< A2> … Gn=base<An>.

### 3.4 Basic Concepts OF Rough SET THEORY

Rough set theory, an extension of set theory introduced by Pawlak in the late 70's provides a sound basis for the study of information system to extract the vital information hidden in the data set. The main idea of rough approximation is that a set can be represented by a lower approximation and an upper one. One of the objectives of rough set data analysis is to reduce data size by finding the reduct, the minimal set of attributes which preserves the classification power of the original set. The main advantage of the rough set data analysis is that it does not use information outside the target data sets.

### 3.5 Information System

An information system is defined in [7], as a set of objects S = { U, A, V, f } where U set of objects, A , attributes ,V , the value given to the attributes defined by the, relation f as f , U X A → V

### 3.6 Equivalence relation

Let X be a set and let x, y, and z be elements of X. An equivalence relation R on X is a Relation on X such that R satisfies (i)Reflexive Property, xRx for all x in X (ii) Symmetric Property, if xRy, then yRx and (iii) Transitive Property, if xRy and yRz, then xRz.

### 3.7 Equivalence classes

The equivalence relation R defines a partition on A; such a partition is a set of all equivalence classes of R.

### 3.8 Discernibility and Indiscernibility

Discernibility and Indiscernibility relation of objects in the information system in rough set theory are two main concepts which are very effective in classification, characterization and clustering the objects. Discernibility matrix consists of entries or set of attributes which discern two objects, where the indiscerniblity matrix IDxy consist entries of attributes which are common to two objects defined by

IDxy = { a ∈ A , f (a, x) = f(a, y)} where x, y ∈ U }

### 3.9 Reduct

Reduct is the minimal set of attributes preserving classification power on original data set A [9], which can be derived by finding the basis of the vector space.

## 4 METHODOLOGIES

This section is concerned with a viable and effective method for the computation of reduct. The idea of the topological concept is used in [10], to find the reducts by find-ing the base for the equivalence relation set taken from the given information system. The minor disadvantages of this process are, (1) if the data is very large, the process to find the base by removing the set of elements corresponding to the attributes one by one without any measure on the significance of attributes randomly makes the process uncertain and (2) it will not prevent unnecessary computations. In order to avoid unnecessary computations, a modified procedure has been suggested. In the first phase, instead of checking the redundancy of attributes in a random manner as proposed in [18], attributes are checked in the descending order according to their measure of significance which guides in the promising direction and in turn reduces the uncertainty and saves the computational costs which is more important in process of feature filtering. This measure may not be fully sufficient to decide the most significant feature. Anyhow, after testing a few databases this measure shows that definitely the first attribute having the least measure of the significance and atleast a few attributes in the initial stages are most probably redundant. This will make the process of elimination of redundant attributes in the initial stage itself which in turn reduce the unnecessary computations. In the second phase, comparing the base after removing features corresponding to the most probable redundant attribute, with the original sub base, will confirm the redundancy of the attribute. Another primary challenge in data mining is handling a large set of data. To meet the challenge of mining from a vast amounts of data, a viable method using algebraic topological techniques is proposed for the computation of the base from the sub base as a base of union of bases of its disjoint subsets which makes the process further simple by using the algebraic topological property, base <A> = base< A1 U A2 > = base< base<A1> U base<A2> >, where A=A1UA2. Based on the methodology described here two algorithms have been developed. Algorithm1 describes the method to find the base from a sub base using indiscernity matrix. Algorithm 2 describes the method to compute the base from the base of union of two bases which will greatly reduce the complexity of the process since large data set is divided into a union of small data sets.

## Algorithm 1: To compute the base (reduct) from indiscernity matrix

Step 1: Convert the information system I into a matrix form.

Step 2: Equivalence relation set A is formed from the given information system I.

Step 3: The equivalence relation set A got in step 2 is converted into a matrix B.

Step 4: Find the indiscernity matrix for the matrix B

Step 5: Remove the repeated indiscern elements from the indiscernity matrix of B

Step 6: Again find the indiscernity matrix from the remaining set of indiscernable elements.

Step 7: Continue this process until there is no further indiscernable elements.

Step 8: If all the elements of the given set X are not available in the last indiscernity matrix, all missing elements are taken from the previous indiscernity matrices which

will form the base G for the topology $\Im$ on X

## Algorithm 2: To compute the base (reduct) by dividing the entire data set into subsets

Step 1: Convert the information system I into a matrix form.

Step 2: Divide the entire information system A into two disjoint sets A1 and A2

Step 3: Find the base of A1 and base of A2 seperately.

Step 4: Compute base of A from the base of union of bases of A1 and A2.

Base<A>=base < base <A1> U base <A2> >

## Algorithm 3: To compute the base (reduct) after eliminating the redundant attributes

Step 1: Convert the information system I into a matrix form.

Step 2: Compute measure of significance for all the attributes in the information system.

Step 3: Attributes in A are arranged according to the ascending order of their measure of significance.

Step 4: Sub base A is divided into two disjoint sets A1, A2 where the first set A1 has members corresponding to the attributes having minimum measure of significance (number of elements in A1 and A2 can be decided with respect to the problem) and A2 has members corresponding to the attributes having next measure of significance.

Step5: The attribute which have least measure of significance value is first removed from A1 and if base<A>=base<base<(A1-(members corresponding to the removed attribute))>U base<A2>> confirms the redundancy of that removed attribute.

Step 6: The process is repeated until all attributes in set A1 are tested for their redundancy and the same process is repeated for A2 also.

Step 7: The set of all attributes which cannot be removed from A1 and A2 is the reduct of the attributes of the given information system.

## Illustration

Algorithms 1, 2 and 3 are illustrated with an example discussed in [10], of seven segments display of the numbers.

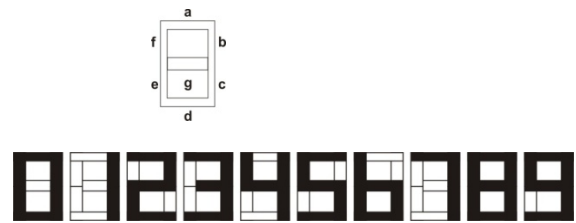

Fig.1 seven segments display of the numbers

## 4.1 Illustration of Algorithm 1 with an example

Algorithm 1: To find the base from a sub base using indiscernity matrix is illustrated with an example of seven segments display of the numbers.

| U\A | a | b | c | d | e | f | g |
|---|---|---|---|---|---|---|---|
| 0 | 0 | 1 | 1 | 1 | 1 | 1 | 0 |
| 1 | 1 | 0 | 1 | 1 | 0 | 0 | 0 |
| 2 | 1 | 1 | 0 | 1 | 1 | 0 | 1 |
| 3 | 1 | 1 | 1 | 1 | 0 | 0 | 1 |
| 4 | 0 | 1 | 1 | 0 | 1 | 1 | 1 |
| 5 | 1 | 0 | 1 | 1 | 0 | 1 | 1 |
| 6 | 1 | 1 | 0 | 1 | 1 | 1 | 1 |
| 7 | 1 | 1 | 1 | 1 | 0 | 0 | 0 |
| 8 | 1 | 1 | 1 | 1 | 1 | 1 | 1 |
| 9 | 1 | 1 | 1 | 1 | 1 | 0 | 1 |

Step 1: As a ... rix re-
presentatio... umbers
0,1,2,3,4,5,6,... given in the ... ure 1 is
given in the ...

| | 1 | 2 | 3 | 4 | 5 | 6 | 7 | 8 | 9 | 10 | 11 | 12 | 13 | 14 |
|---|---|---|---|---|---|---|---|---|---|---|---|---|---|---|
| 1 | 1 | | 0,2,3,7,8,9 | 5,6 | 0,3,5,6,7,8,9 | 7 | 0,2,3,5,6,8,9 | 7 | 0,2,3,5,6,8,9 | 3,5,7,9 | 0,5,6,8,9 | 2,3,7 | 0,7 | 2,3,5,6,8,9 |
| 2 | | | 1,4 | | 1,4 | | | 1,4 | | 1,4 | 4 | 1 | 1 | 4 |
| 3 | | | | 0,1,3,4,7,8,9 | 0,2,3,8,9 | | 0,2,8 | 1,4,7 | 0,4,8,7,9 | 1,9 | 0,1,2,3,7 | 1,3,7 | 0,1,8,9 | 2,3,4,8,9 |
| 4 | | | | 1 | 5,6 | nil | 5,6 | nil | 6 | | 5,6 | nil | nil | 5,6 |
| 5 | | | | | 1 | | 0,3,5,6,8,9 | 1,4,7 | 0,6,8 | 1,3,4,5,7,8,9 | 0,4,5,6,8,9 | 1,3,7 | 0,3,4,5,6,8,9 |
| 6 | | | | | | 1 | | nil | 2 | nil | nil | 2 | nil | 2 |
| 7 | | | | | | | 1 | | 0,2,6,8 | 3,5,9 | 0,5,6,8,9 | 2,3 | 0 | 2,3,5,6,8,9 |
| 8 | | | | | | | | 1 | | 1,4,7 | 4 | 1,2,7 | 1,2 | 2,3 |
| 9 | | | | | | | | | 1 | nil | 0,6,8 | 2 | 0 | 2,6,8 |
| 10 | | | | | | | | | | 1 | 4,5,9 | 1,5,7 | 1,3,5,7,9 | 3,4,5,9 |
| 11 | | | | | | | | | | | 1 | nil | 0 | 4,5,6,8,9 |
| 12 | | | | | | | | | | | | 1 | 1,2 | 2,3 |
| 13 | | | | | | | | | | | | | 1 | nil |
| 14 | | | | | | | | | | | | | | 1 |

Step 2: Equivalence relation set A of the information table1 given in step 1 which is taken as a sub base of the information system is given below

A= {U/IND(a) = { 0,2,3,5,6,7,8,9}, {1,4},

U/IND(b)= {0,1,2,3,4,7,8,9}, {5,6},

U/IND(c)= {0, 1, 3,4,5,6,7,8,9}, {2},

U/IND(d)= {0, 2,3,5,6,8,9}, {1,4,7},

U/IND(e)= {0, 2, 6, 8},{1, 3, 4, 5, 7, 9},

U/IND(f)= {0, 4,5,6,8,9},{1, 2, 3,7},

U/IND(g)= {0, 1, 7} ,{2, 3, 4, 5, 6, 8, 9} }

Step 3: Equivalence relation matrix B formed from the equivalence relation set A from which indiscernity matrix to be found is given in Table 2.

**Table 2**
**Equivalence relation matrix B**

| | 0 | 1 | 2 | 3 | 4 | 5 | 6 | 7 | 8 | 9 |
|---|---|---|---|---|---|---|---|---|---|---|
| 1 | 0 | - | 2 | 3 | - | 5 | 6 | 7 | 8 | 9 |
| 2 | - | 1 | - | - | 4 | - | - | - | - | - |
| 3 | 0 | 1 | 2 | 3 | 4 | - | - | 7 | 8 | 9 |
| 4 | - | - | - | - | - | 5 | 6 | - | - | - |
| 5 | 0 | 1 | 2 | 3 | 4 | 5 | 6 | 7 | 8 | 9 |
| 6 | - | - | - | - | - | - | - | - | - | - |
| 7 | 0 | 1 | 2 | 3 | - | 5 | 6 | - | 8 | 9 |
| 8 | - | 1 | - | - | 4 | - | - | 7 | - | - |
| 9 | 0 | 1 | 2 | 3 | - | - | 6 | - | 8 | - |
| 10 | - | 1 | - | 3 | 4 | - | 6 | 7 | - | 9 |
| 11 | 0 | - | - | - | 4 | 5 | 6 | - | 8 | 9 |
| 12 | - | 1 | 2 | 3 | - | - | - | 7 | - | - |
| 13 | 0 | 1 | - | - | - | - | - | 7 | - | - |
| 14 | 0 | 1 | 2 | 3 | 4 | 5 | 6 | - | 8 | 9 |

Steps 4, 5, 6 and 7: The results of the process explained in algorithm1 in steps 4, 5 and 6 to find the indiscernible elements, until there is no indiscernible elements could be found from the indiscernity matrix.

| | | | | | | | | | | | | |
|---|---|---|---|---|---|---|---|---|---|---|---|---|
| Iteration 5 | {0} | {0,8} | {1} | {2} {3} {3,9} | {4} {5} {5,6} | {5,9} | {6} | {6,8} | {7} | {8} | {8,9} | {9} |
| Iteration 6 | {0} | {3} | {5} | {6} {8} {9} | - | - | - | - | - | - | - | - |

Base for the entire set X=G={0,3,5,6,8,9 U members corresponding to the missing elements taken from the previous indiscernity matrix }. Therefore the sets containing the missing terms 1,2,4,7 are taken from the previous indiscernity matrix and therefore the base for the topology T on X =

G={{0},{3},{5},{6},{8},{9},{1},{2},{4},{7}}

## 4.2 Illustration of Algorithm 2, 3 with an example

Algorithm 2: To find the base from a sub base using indiscernity matrix is illustrated with an example of seven segments display of the numbers.

Step 1: As an initial step of the algorithm, the matrix representation of seven segments display of the numbers 0,1,2,3,4,5,6,7,8,9 depicted in figure 1 is given in information table1 as described in [10].

Steps 2, 3: The results of the measure of significance of attributes [22] are calculated difference between the positive region of the decision rules with and without the attribute concerned using the data mining software RSES (Rough Set Exploration System) are given in Table 3.

From the measure of the significance of all the attributes given in ascending order in Table 3 it is found that the most probable redundant attributes are identified based on the significance of attributes. In this example the attributes c, d have the least and attribute a, f, g have the next minimum measure.

**TABLE 3**
**Significance of Attributes**

| Sl.No. | Attribute | Measure of significance |
|---|---|---|
| 1 | C | 0 |
| 2 | D | 0 |
| 3 | A | 0.2 |
| 4 | F | 0.2 |
| 5 | G | 0.2 |
| 6 | B | 0.4 |
| 7 | E | 0.4 |

Step 4: The small disadvantage of the method proposed in [18] which is explained in step3 is rectified in step 4 by dividing the sub base into union of sub sets based on measure of significance and applying algebraic topological techniques in the information system. We divide the set A, the set of equivalence relations on X into two sets A1and A2 with lower and higher measure of significance respectively

A1 = { U/IND(c) = {0, 1,3,4,5,6,7,8,9} {2},
    U/IND(d) = {0,2,3,5,6,8,9} {1,4,7}
    U/IND(a) = { 0,2,3,5,6,7,8,9} {1,4},
    U/IND (f) = {0, 4, 5, 6, 8, 9} {1, 2, 3, 7}
    U/IND (g) = {0, 1, 7} {2, 3, 4, 5, 6, 8, 9} }

A2 = {U/IND (b) = {0, 1, 2, 3, 4, 7, 8, 9} {5, 6},
    U/IND (e) = {0, 2, 6, 8} {1, 3, 4, 5, 7, 9} }

Step 5: Form a new set B1 by removing the elements from A1 corresponding to the attribute having least significant measure (in this example attribute 'c').
B1={ {0,2,3,5,6,8,9} {1,4,7}
{0, 2, 3, 5, 6, 7, 8, 9}{1,4}, {0, 4, 5,6,8,9} {1, 2, 3,7}, {0, 1, 7} {2, 3, 4, 5, 6, 8, 9} }
A2 ={{ {0, 1,2,3,4,7,8,9} {5,6}, {0, 2, 6, 8} {1, 3, 4, 5, 7, 9} }
base<B1>=G1={{0},{1}{2,3},{4},{7}, {5,6,8,9} }
base < A2>= G2={{0,2,8},{5},{6},{1,3,4,7,9}}
base<A>= base< B1 U A2 > = base< base<B1> U base<A2> >
base<A>=base<G1UG2>={{1},{2},{3},{4},{5},{6},{7},{8},{9}, {10} }
which is same as the base for the entire set X which proves that attribute 'c' is a redundant attribute. Therefore 'c' will leave the basis.
Steps 6 and 7: The process done in step 5 is repeated for all the remaining attributes in the descending of its significance values, in the set B1 and A2. By proceeding like this, it is proved that attribute'd' is also a redundant attribute. Therefore attribute'd' also left the basis. The set of remaining attributes {a, e, b, f, g} got in the final step form the reduct of the given of information system.

## 5 CONCLUSIONS

This paper has dealt with the challenging data mining problem of feature reduction to extract the hidden information from a huge set of data in a simplified method. In the method proposed in this work to find the reduct-a minimal set of features that preserves the indiscernibility relation of an information system, it is found that the elimination of the redundant attributes based on measure of significance of attributes guides in the promising direction which in turn reduces the uncertainty and saves the computational costs which is more important in feature filtering process instead of going in a random manner. Linear correlation measure between attributes will be effective only if the features are linearly correlated, but linear correlation cannot always be expected between the features. The concept of measure of significance of attributes used in this study overcomes this shortcoming. Also the method proposed in this paper for the computation of the base from the sub base as a base of union of bases of its disjoint subsets applying algebraic topological property will reduce the challenging task of handling a large set of data at a time. Another advantage of this method is that it will derive the most required reduct of the information system. The algorithm proposed in this work can be extended to a problem with uncertain and missing data. It is intended to expand the model proposed in this study to improve further the certainty in the filtering process of redundant attributes using weighted rough set model.

**P.G.JansiRani -** she completed her MSc (Applied Mathematics) in 1987 and M.Phil (Mathematics)in the year 1989, Bharathidasan University,Tiruchirapalli . She worked as lecturer in the Department of Mathematics at periyar Maniyammai College of Technology for women, Tamil Nadu from 1990 to 1995.Since 1995 she is working as Professor in the Department of Mathematics at Sethu Institute of Technology, Virudhunagar, Tamilnadu. She has 20 years of teaching experience in Engineering College. She is also a research scholar at School of Mathematics, Madurai Kamaraj University, Madurai, Tamilnadu, India. Fields of interest: Data Mining and Rough set theory, Operations Research.

**R.Bhaskaran** - He did his M.Sc., at IIT Chennai in 1973 and obtained his Ph.D at the Ramanujan Institute for Advanced study in Mathematics, University of Madras in 1980. He joined the Madurai Kamaraj University in 1980, as a lecturer. Now, he is working as a senior professor in the School of Mathematics. His areas of interest are Non-archimedean functional analysis; Image processing, Data Mining, Software development for learning Mathematics and Character Recognition.